\documentclass{appolb}
\usepackage{graphicx}
\usepackage{lineno}
\usepackage{xspace}
\usepackage{amsmath}
\def\Pp      {\ensuremath{p}\xspace} 
\def\proton      {{\ensuremath{\Pp}}\xspace}
\def\pp      {{\ensuremath{\proton\proton}}\xspace}
\def\pPb   {\ensuremath{\proton\mathrm{Pb}}\xspace}
\def\Pbp   {\ensuremath{\mathrm{Pb}\proton}\xspace}
\def\c      {\ensuremath{\mathrm{c}}\xspace}  
\def\j      {\ensuremath{\mathrm{j}}\xspace}  
\def\R      {\ensuremath{\mathrm{R}}\xspace}    
\def\N      {\ensuremath{\mathrm{N}}\xspace} 
\def\PZ      {\ensuremath{\mathrm{Z}}\xspace}  
\def\Z      {{\ensuremath{\PZ}}\xspace}
\def\Zc      {\ensuremath{\PZ\c}\xspace}  
\def\Zj      {\ensuremath{\PZ\j}\xspace}  
\def\to     {\ensuremath{\rightarrow}\xspace} 
\def\Pmu    {\ensuremath{\mu}\xspace}
\def\mumu   {{\ensuremath{\Pmu^+\Pmu^-}}\xspace}
\def\Zmumu   {\ensuremath{\Z\to\mumu}\xspace}

\newcommand{\aunit}[1]{\ensuremath{\text{\,#1}}} 
\newcommand{\tev}{\aunit{Te\kern -0.1em V}\xspace}
\newcommand{\gev}{\aunit{Ge\kern -0.1em V}\xspace}
\newcommand{\mev}{\aunit{Me\kern -0.1em V}\xspace}
\newcommand{\stat}{\aunit{(stat)}\xspace}
\newcommand{\syst}{\aunit{(syst)}\xspace}
\newcommand{\lumi}{\aunit{(lumi)}\xspace}
\newcommand{\lum} {\ensuremath{\mathcal{L}}\xspace}

\def\nb {\aunit{nb}\xspace}
\def\invnb {\ensuremath{\nb^{\mbox{-}1}}\xspace}

\def\fb   {\ensuremath{\aunit{fb}}\xspace}
\def\invfb   {\ensuremath{\fb^{\mbox{-}1}}\xspace}

\def\rfb   {\ensuremath{R_\mathrm{FB}}\xspace}
\def\rpa   {\ensuremath{R_{\proton \mathrm{Pb}}}\xspace}
\def\pT     {\ensuremath{p_{\mathrm{T}}}\xspace}
\def\ZpT     {\ensuremath{p_{\mathrm{T}}^{\Z}}\xspace}
\def\zrap   {\ensuremath{y_{\Z}}\xspace}
\def\zrapstar {\ensuremath{y^{*}_{\Z}}\xspace}

\begin{document}
\title{Probing the valence quark region of nucleons with  Z  bosons at LHCb%
\thanks{Presented at 2022 Quark Matter conference}%
}
\author{Tianqi Li, on behalf of the LHCb collaboration 
\address{Guangdong Provincial Key Laboratory of Nuclear Science, Guangdong-Hong Kong Joint Laboratory of Quantum Matter, Institute of Quantum Matter, South China Normal University, Guangzhou, China}
}
\maketitle
\begin{abstract}
Because of its forward coverage, LHCb can probe the valence quark distributions of protons and nuclei at small and large Bjorken-x with high precision.
This proceeding presents new LHCb measurements of Z boson production in association with charm jet in the forward region of proton-proton collisions and Z boson production in proton-lead collisions.
Z+charm jet production could be sensitive to a valence-like intrinsic-charm component in the proton wave function.
The measurements of Z production in proton-lead collisions provide new constraints on the partonic structure of nucleons inside nuclei. Comparisons between the results and calculations with parton distribution functions are also discussed.
\end{abstract}
  
\section{Introduction}
\begin{figure}[htb]
\centerline{
\includegraphics[width=0.7\linewidth]{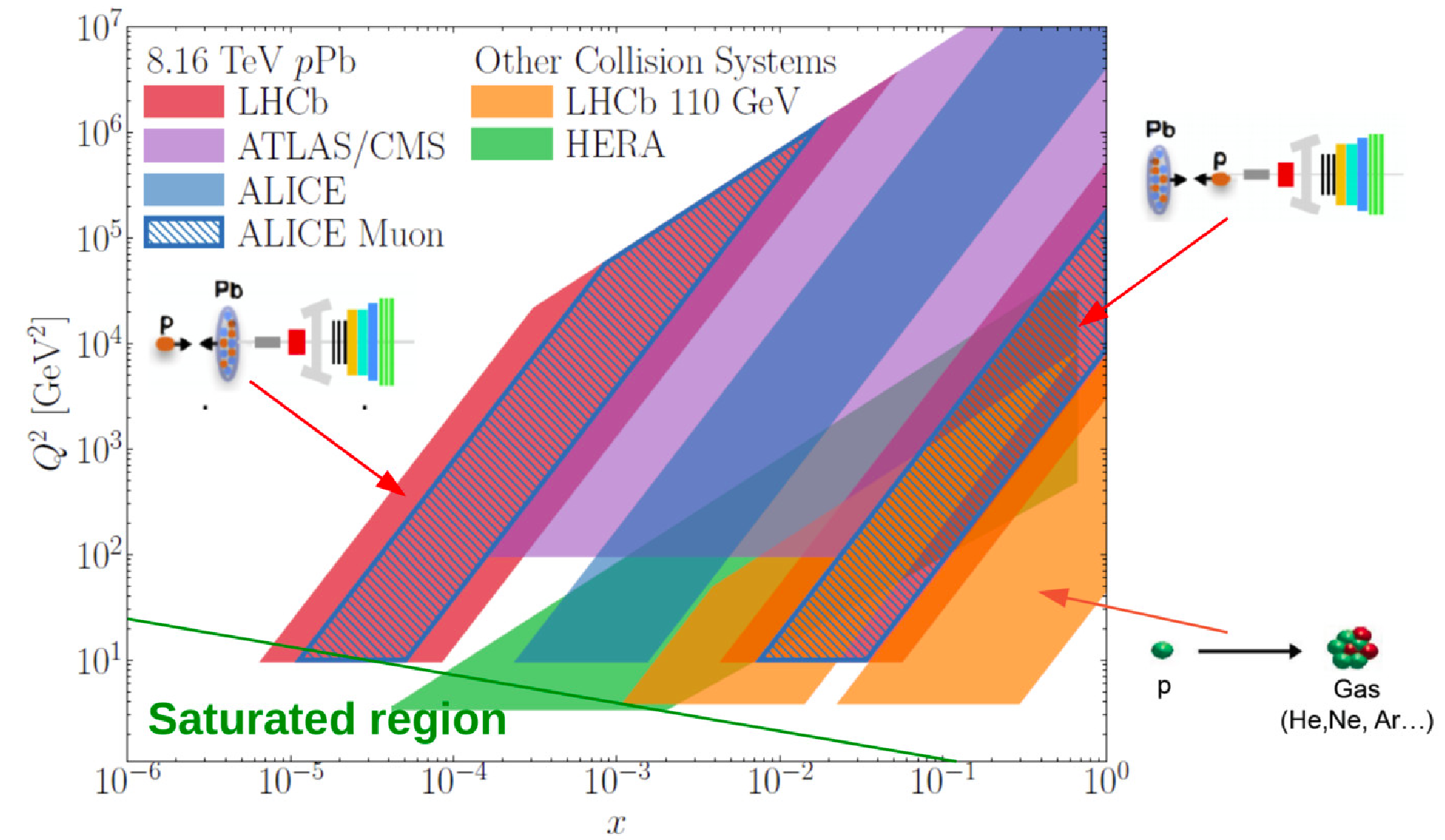}}
\caption{Kinematic regions corresponding to the acceptance of the four LHC experiments in
p-Pb collisions.}
\label{fig:lhcb-phase-space}
\end{figure}

The LHCb detector~\cite{1.1, 1.2}, optimized for studying heavy flavour physics in proton-proton (\pp) collisions at the LHC,
is a single-arm forward spectrometer covering the pseudorapidity range $2 < \eta < 5$, 
providing a high momentum resolution down to very low transverse momentum (\pT), excellent reconstruction and particle identification.
Due to the forward acceptance, LHCb results are highly complementary to the other LHC experiments.
LHCb participated successfully in the recording of proton-lead data samples in 2013, thus becoming a member of the LHC heavy ion program to provide unique datasets for the heavy ion physics studies.
As shown in Fig.~\ref{fig:lhcb-phase-space}, LHCb provides data that can constrain nuclear parton
distribution functions (nPDFs) in both low-x and high-x regions.
The measurements of Z boson production at LHCb can provide constraints at high momentum transfer ($Q^2$).
Furthermore, the LHCb measurements of Z+charm jet production provide information on the quark PDFs at large-x region relevant to studies of intrinsic charm.

In this article, we report on recent results of the Z bosons produced in association with charm in the forward region~\cite{1.3}
and the Z boson production in the proton-lead collisions~\cite{1.4} measured by the LHCb collaboration.

\section{Probe intrinsic charm: \Zc production in proton-proton collisions}

The possibility that the proton wave function may contain an intrinsic charm (IC) component, in addition to the c-quark content due to perturbative gluon radiation, has been an interesting topic in recent years (see Ref.~\cite{2.1} for a review). 
Light front QCD calculations (Ref.~\cite{2.6,2.7}) predict the presence of the non-perturbative IC can manifest in valence-like charm content in the parton distribution function, whereas, if the charm is entirely perturbative, the charm PDF should rapidly decrease at large Bjorken-x.
The goal is to use a high-precision measurement of an observable with direct sensitivity to large-x charm to probe IC in proton, where Q is large enough that hadronic and nuclear
effects are negligible.

\begin{figure}[hbpt]
  \begin{center}
    \includegraphics[width=0.49\linewidth]{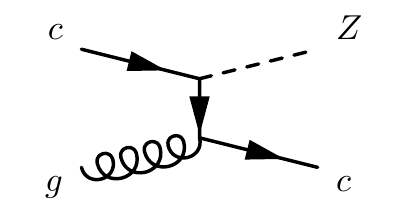}
    \put(-170,70){(a)}
    \includegraphics[width=0.49\linewidth]{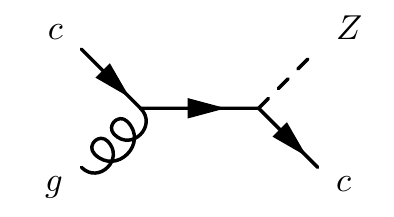}
    \put(-170,70){(b)}
    \vspace*{-0.5cm}
  \end{center}
\caption{Leading-order Feynman diagrams for gc$\rightarrow$\Zc.}
  \label{fig:Feynman-diagrams}
\end{figure}

Ref.~\cite{2.2} proposes probing IC by studying events containing a \Z boson and a charm jet, \Zc, in the forward region of \pp collisions at the LHC. Since \Zc production is inherently at large Q, and the electroweak bosons do not participate in the strong interaction, hadronic effects are small.
Fig.~\ref{fig:Feynman-diagrams} shows gc\to \Zc scattering, a leading-order \Zc production mechanism where one of the initial partons should have large x in the forward rapidity region. Thus \Zc production serves as a direct probe of IC in valence quark region. 

This article presents the first measurement of \Zc production in the forward region using the full Run 2 data collected by LHCb in 2015, 2016, 2017, and 2018 at 13 \tev and corresponding to 6 \invfb.
The strategy of this analysis is to measure the ratio of production cross sections $\R^c_j \equiv \sigma(\Zc)/\sigma(\Zj)$, where \Zj refers to events containing a Z boson and any type of jet. The ratio is chosen because it is less sensitive to experimental and theoretical uncertainties than $\sigma(\Zc)$. 
The \Z bosons are reconstructed using the \Zmumu decay in the mass range $60 < m(\mu^+\mu^-) < 120 \gev$. 
The quantity $\R^c_j$ is measured as $\R^c_j = \N(c\mbox{-}\text{tag})/[\epsilon_{c\mbox{-}\text{tag}}\N(j)]$, where $\N(c\mbox{-}\text{tag})$ is the observed \Zc yield, $\epsilon_{c\mbox{-}\text{tag}}$ is the c-tagging efficiency, and $\N(j)$ is
the total \Zj yield.

\begin{figure}[htb]
\centerline{
\includegraphics[width=0.7\linewidth]{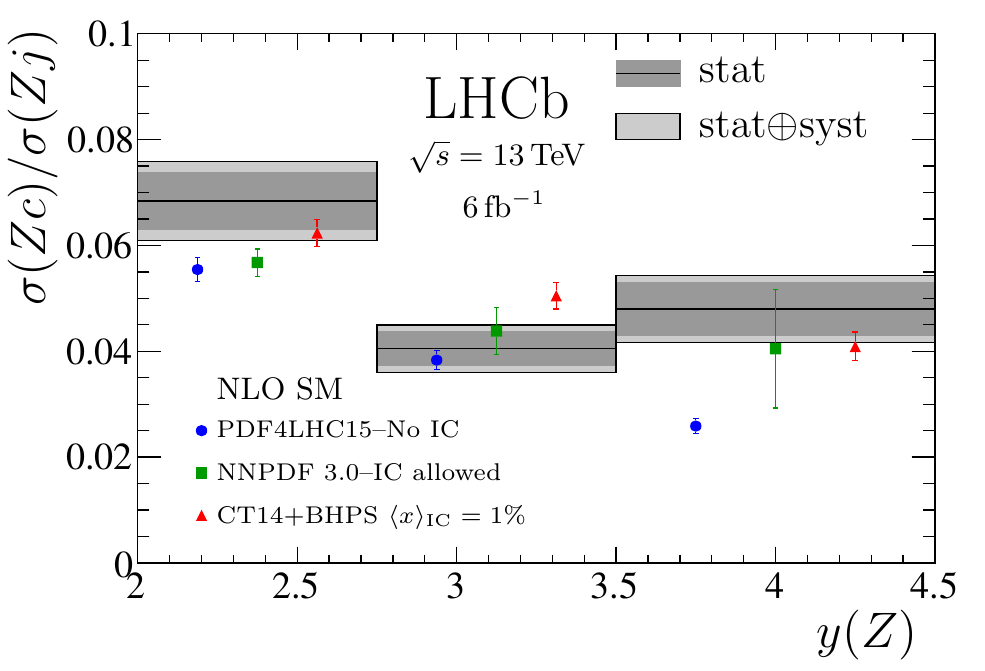}}
\caption{Measured $\R_j^c$ distribution compared to NLO SM predictions without IC, with the c quark PDF shape allowed to vary, and with IC as predicted by BHPS with a mean momentum fraction of 1\%.}
\label{fig:Fig5}
\end{figure}

Fig.~\ref{fig:Fig5} shows the measured $\R^c_j$ distribution in bins of \Z rapidity (\zrap). The ratio is determined to be $4.98\pm0.25\stat\pm0.35\syst$ in the range of $2.0<\zrap<4.5$.
NLO SM predictions for $\R^c_j$ without IC~\cite{2.3}, allowing for potential IC~\cite{2.4}, and with the valence-like IC predicted by BHPS with a mean momentum fraction of $1\%$~\cite{2.5} are also given and compared with the measurements. 
The observed ratios are consistent with both the no-IC and IC hypotheses in the first two \zrap bins, whereas it is not the case in the most forward bin where valence-like IC is expected to cause a large enhancement.
The observed $\R_j^c$ value is significantly larger than the no-IC-expected calculation, and is consistent with the calculations including the IC.

\section{Probe cold nuclear matter effects: \Z production in proton-lead collisions}
The production of electroweak bosons at hadron colliders is an important benchmark process. 
Because electroweak bosons do not participate in the strong interaction, they are not affected by the QCD medium created in proton-lead collisions. 
Therefore, electroweak bosons preserve the initial conditions of the collisions, 
and can be used to probe cold nuclear matter effects and constrain nPDFs~\cite{3.1,3.2} inside nuclei.

This article presents the measurements of the \Zmumu production in proton-lead collisions at LHCb using data samples collected at a center-of-mass energy of $8.16 \tev$. 
The dataset corresponds to an integrated luminosity of $12.2 \pm 0.3 \invnb$ for forward (\pPb) collisions and $18.6 \pm 0.5 \invnb$ for backward (\Pbp) collisions.
The fiducial cross-section of the \Zmumu production is measured using the formula $\sigma_{\Zmumu}=(\N_{\text{cand.}}\cdot\rho\cdot f_{\text{FSR}})/(\lum\cdot\epsilon_{\text{tot}})$, 
where $\N_{\text{cand.}}$ is the number of \Zmumu candidates after signal selection, 
$\rho$ is the signal purity, 
$f_\text{FSR}$ is final state radiation correction,
$\lum$ is the integrated luminosity, 
and $\epsilon_{\text{tot}}$ is the total efficiency.
The fiducial region is defined as $60 < m_{\mumu} < 120 \gev$ where both muons satisfy $\pT > 20 \gev$ and $2 < \eta < 4.5$.
The total fiducial cross-section of the Z production in forward and backward collisions is measured to be
$\sigma_{\Zmumu,\,\pPb}= 26.9\pm 1.6\stat\pm 0.9\syst\pm 0.7\lumi \nb $ and $\sigma_{\Zmumu,\,\Pbp}= 13.4\pm 1.0\stat\pm0.5\syst\pm0.3\lumi \nb$. 

The forward-to-backward production ratio, \rfb, is sensitive to nuclear effects in the \Zmumu production.
The \rfb ratio in the common Z boson rapidity acceptance window is measured to be $0.78\pm0.10$.
Comparing the cross-sections between proton-lead and pp collisions, the nuclear modification factors (\rpa) are also calculated. The measured nuclear modification factors are $0.94\pm0.07$ and $1.20\pm0.11$ for forward and backward rapidities, respectively. 

\begin{figure}[htbp]
	\begin{center}
		\includegraphics[width=0.49\linewidth]{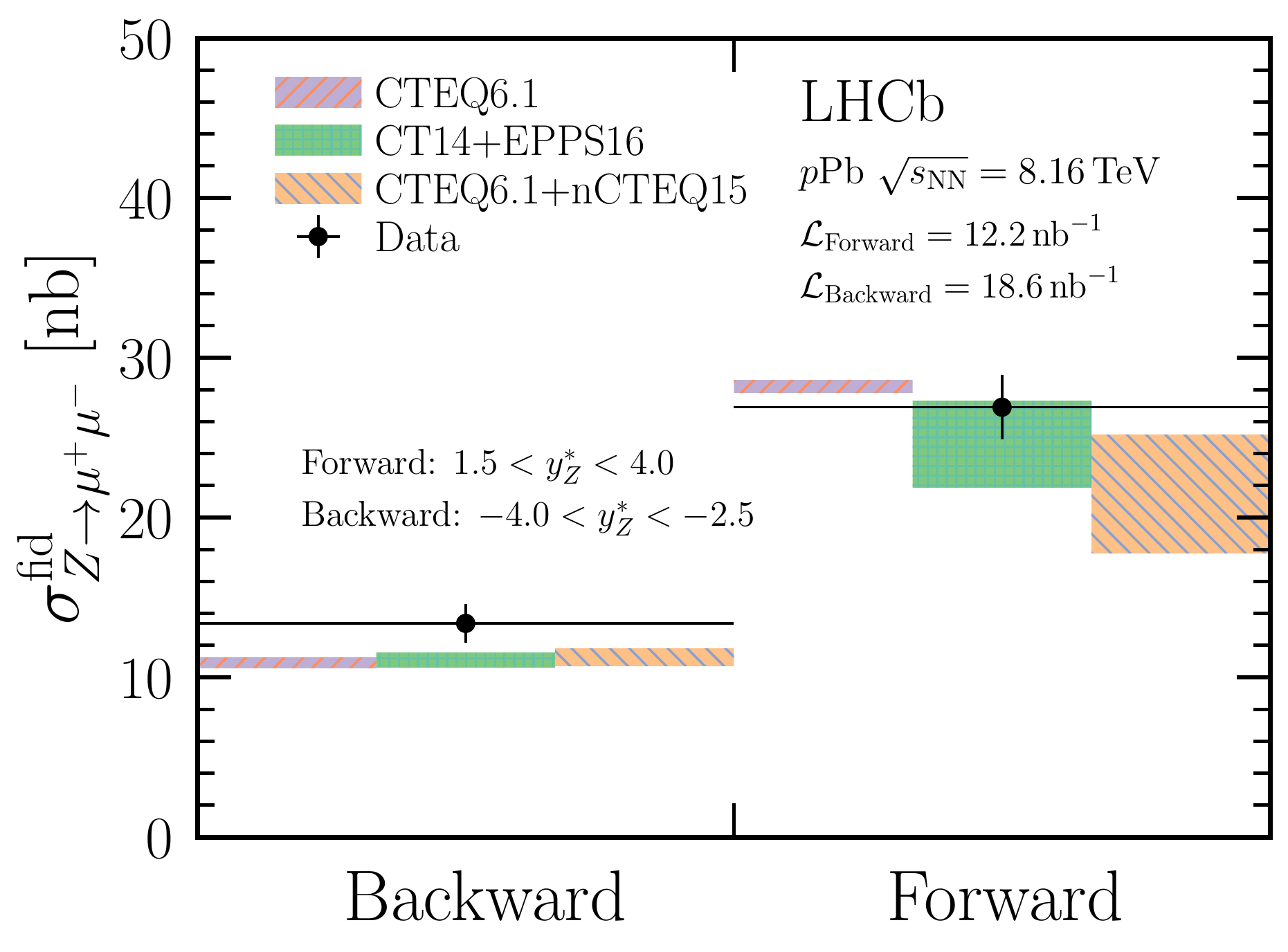}
		\includegraphics[width=0.49\linewidth]{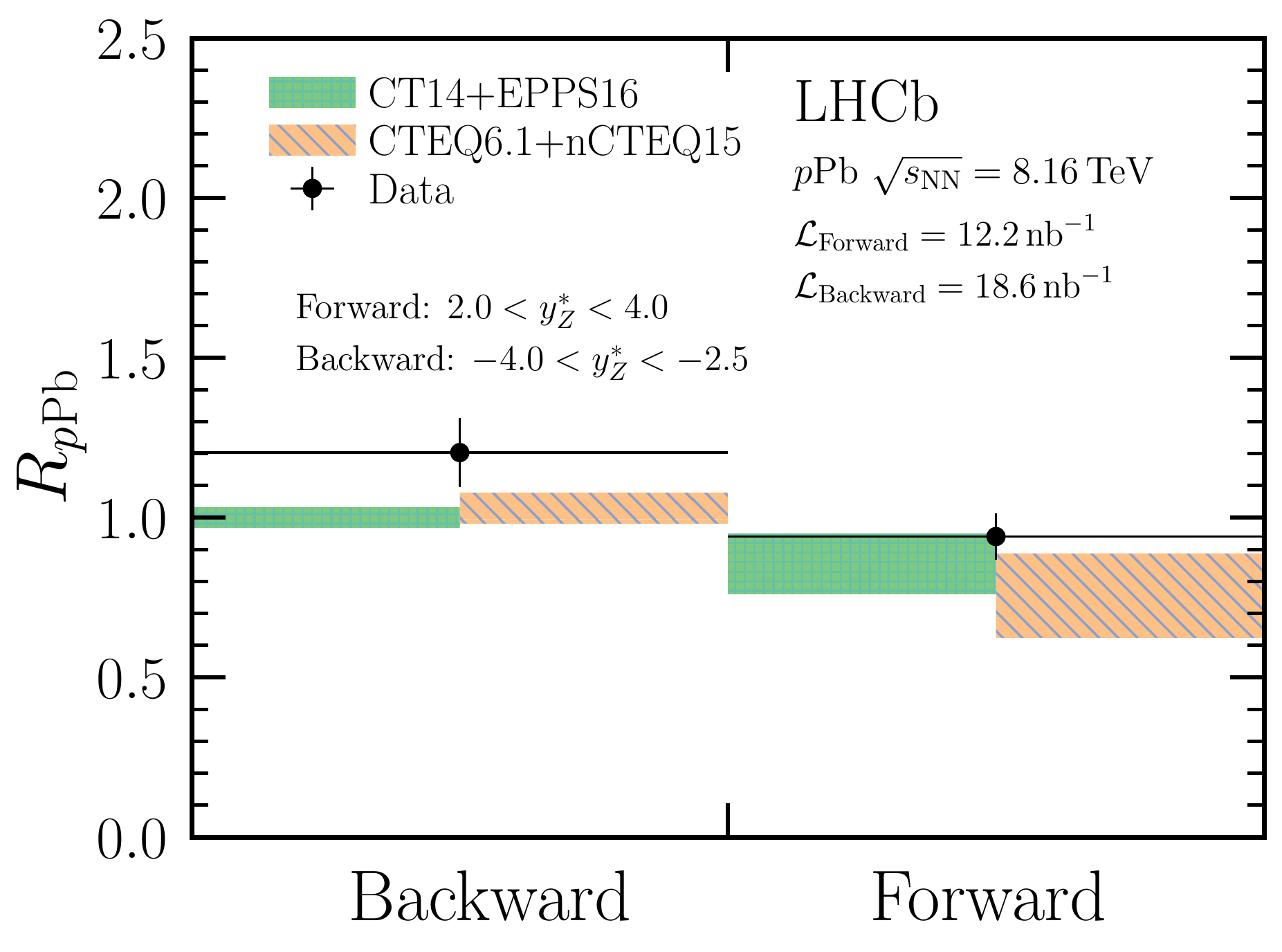}
		\includegraphics[width=0.50\linewidth]{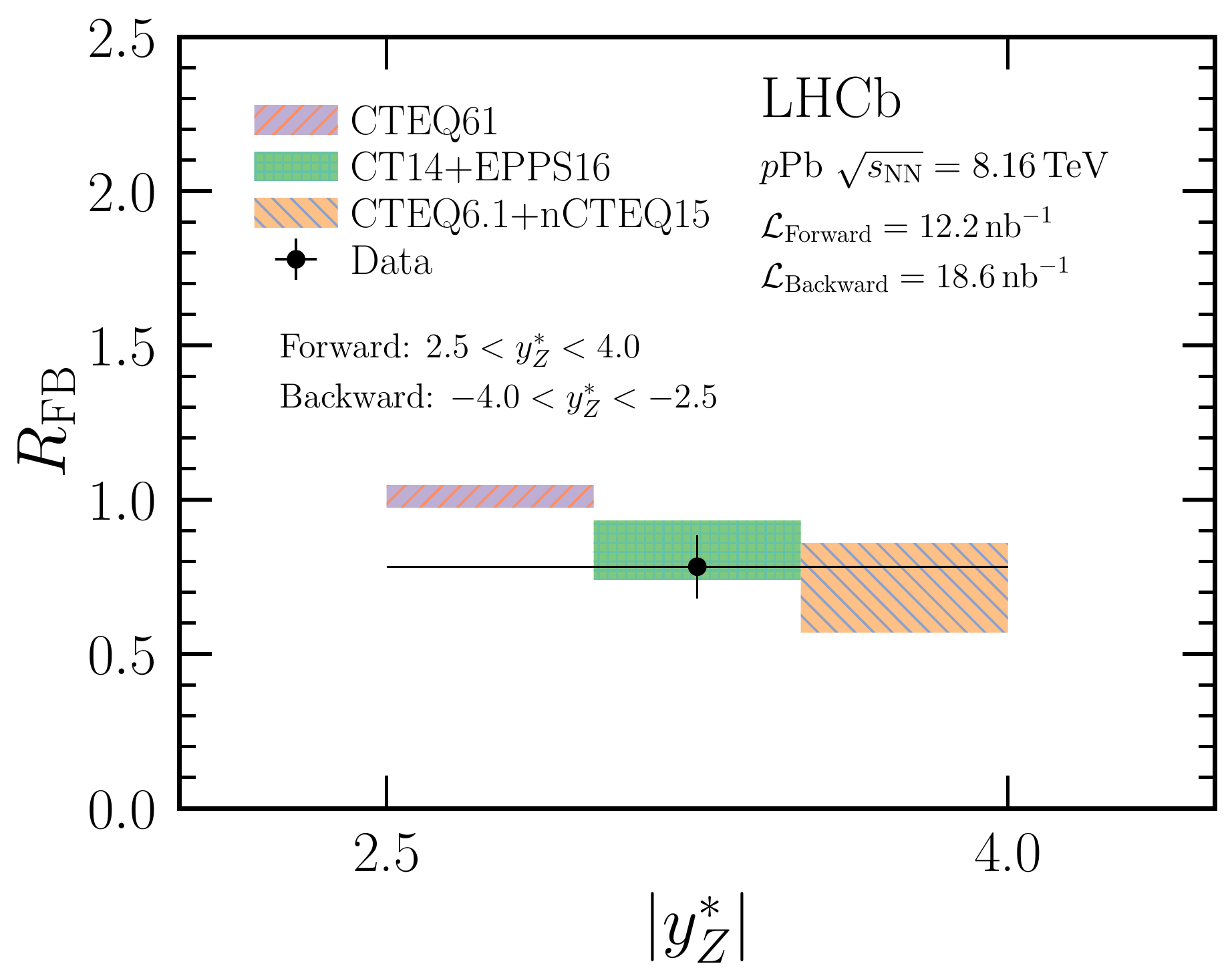}
	\end{center}
	\caption{
		The measured overall Z production fiducial cross-section (top-left), the measured nuclear modification factor (top-right) and the measured forward-backward ratio (bottom), compared with the PowhegBox prediction using CTEQ61, EPPS16 and nCTEQ15 (n)PDF sets, for forward and backward collisions, respectively.}
	\label{fig:xsec}
\end{figure}

Fig.~\ref{fig:xsec} shows the measured Z boson cross-section, \rfb and \rpa, compared to the POWHEG~\cite{3.3,3.4,3.5} calculations of the CTEQ61 proton PDF set, EPPS16 nPDF and nCTEQ15 nPDF sets. The differential results as a function of Z rapidity in the centre-of-mass frame (\zrapstar), the transverse momentum (\ZpT) and an angular variable $\phi^*$ can be found in Ref.~\cite{1.4}. The variable $\phi^*$~\cite{3.6} is defined as $\phi^*=\tan(\pi/2-|\Delta\phi|/2)/\cosh(\Delta\eta/2)$, where the $\Delta\phi$ is difference between the azimuthal angle of the two muon momenta, the $\Delta\eta$ is difference between the pseudorapidity of the two muons.

\section{Conclusions}
In this proceeding, we present the first measurement of events containing Z boson and a charm jet in the forward region of pp collisions at 13 \tev. 
The ratio $\R_j^c$ as a function of \zrap is measured, and is compared to NLO SM calculations. 
The observed $\R_j^c$ shows a sizable enhancement in forward \zrap, consistent with the expectation of an intrinsic charm presence.

Z boson production in \pPb collisions can probe the cold nuclear matter effects. 
This article presents the measurements of the Z boson production in proton-lead collisions at 8.16 \tev at LHCb.
The Z-boson production fiducial cross-section, \rfb and \rpa are measured integrated and differentially as a function of \zrapstar, \ZpT, and $\phi^*$.
These results are compatible with theoretical predictions of the EPPS16 and nCTEQ15 nPDFs. 
The measurements are also compatible with previous results from LHCb experiments at 5.02 \tev~\cite{3.7} but more precise. 
Furthermore, the forward measurements show strong constraining power for the nPDFs, especially at high rapidity region (small Bjorken-x).


\begin{thebibliography}{1}
\bibitem{1.1}
LHCb collaboration, JINST 3 (2008) S08005.
\bibitem{1.2}
LHCb collaboration, Int. J. Mod. Phys. A30 (2015) 1530022.
\bibitem{1.3}
LHCb collaboration, PHYS. REV. LETT. 128 (2022) 082001.
\bibitem{1.4}
LHCb collaboration, ARXIV:2205.10213.
\bibitem{2.1}
S. J. Brodsky et al., Adv. High Energy Phys. 2015 (2015) 231547.
\bibitem{2.6}
S. J. Brodsky et al., Phys. Lett. B93 (1980) 451.
\bibitem{2.7}
S. J. Brodsky et al., Phys. Rev. D23 (1981) 2745.
\bibitem{2.2}
T. Boettcher, P. Ilten, and M. Williams, Phys. Rev. D93 (2016) 074008.
\bibitem{2.3}
T.-J. Hou et al., JHEP 02 (2018) 059.
\bibitem{2.4}
NNPDF collaboration, Eur. Phys. J. C76 (2016) 647.
\bibitem{2.5}
NNPDF collaboration, Eur. Phys. J. C77 (2017), no. 10 663.
\bibitem{3.1}
A. Kusina et al., Eur. Phys. J. C77 (7) (2017) 488.
\bibitem{3.2}
K. J. Eskola et al., Eur. Phys. J. C77 (3) (2017) 163.
\bibitem{3.3}
P. Nason, JHEP 11 (2004) 040.
\bibitem{3.4}
S. Frixione et al., JHEP 11 (2007) 070.
\bibitem{3.5} S. Alioli et al., JHEP 06 (2010) 043.
\bibitem{3.6}
M. Vesterinen and T. R. Wyatt, Nucl. Instrum. Meth. A602 (2009) 432.
\bibitem{3.7}
LHCb collaboration, JHEP 09 (2014) 030.
\end{thebibliography}

\end{document}